\documentclass[10pt,letterpaper]{article}
	\usepackage[top=0.85in,left=2.75in,footskip=0.75in]{geometry}
	\usepackage{amsmath,amssymb}
	\usepackage[utf8x]{inputenc}
	\usepackage{cite}

\usepackage{color}
	\raggedright
\usepackage[aboveskip=1pt,labelfont=bf,labelsep=period,justification=raggedright,singlelinecheck=off]{caption}
	
	\makeatletter
	\renewcommand{\@biblabel}[1]{\quad#1.}
	\makeatother
	\usepackage{lastpage,fancyhdr,graphicx}
	\usepackage{epstopdf}
	\fancyheadoffset[L]{2.25in}
	\fancyfootoffset[L]{2.25in}

\begin{document}
	\vspace*{0.2in}
	\begin{flushleft}
       {\Large
        \textbf\newline{\textbf{Chaotic Attractor Hopping yields Logic Operations}}
        }	
	\newline

	\textbf{K. Murali\textsuperscript{1}, Sudeshna Sinha\textsuperscript{2}, Vivek Kohar\textsuperscript{3,4}, Behnam Kia\textsuperscript{3}, William L. Ditto\textsuperscript{3,*}}

	\bigskip
	
        $^1$Department of Physics, Anna University, Chennai 600 025, India.\\ $^2$ Department of Physical Sciences, Indian Institute of Science  Education and Research Mohali, Knowledge City, SAS Nagar, Sector 81, Manauli PO 140 306, Punjab, India. \\ $^3$Nonlinear Artificial Intelligence Lab, Department of Physics, North Carolina State University, Raleigh, NC 27695, USA.\\$^4$The Jackson Laboratory, 600 Main St, Bar Harbor, ME 04609, USA.

        \bigskip
        
	*wditto@ncsu.edu
	
	\end{flushleft}
	\section*{Abstract}
  Certain nonlinear systems can switch between dynamical attractors
  occupying different regions of phase space, under variation of
  parameters or initial states. In this work we exploit this feature
  to obtain reliable logic operations. With logic output $0/1$ mapped
  to dynamical attractors bounded in distinct regions of phase space,
  and logic inputs encoded by a very small bias parameter, we
  explicitly demonstrate that the system hops consistently in response
  to an external input stream, operating effectively as a reliable
  logic gate. This system offers the advantage that very low-amplitude
  inputs yield highly amplified outputs. Additionally, different
  dynamical variables in the system yield complementary logic
  operations in parallel. Further, we show that in certain parameter
  regions noise aids the reliability of logic operations, and is
  actually necessary for obtaining consistent outputs. This leads us
  to a generalization of the concept of Logical Stochastic Resonance
  to attractors more complex than fixed point states, such as periodic
  or chaotic attractors. Lastly, the results are verified in
  electronic circuit experiments, demonstrating the robustness of the
  phenomena. So we have combined the research
    directions of Chaos Computing and Logical Stochastic Resonance
    here, and this approach has potential to be realized in
    wide-ranging systems.

\bigskip

Keywords: Chaos Computing, Logic gates, Nonlinear Circuits, Logical Stochastic Resonance\\

\medskip

PACS: \ 05.45.-a


\bigskip
\bigskip
\bigskip
\bigskip

\noindent
{\bf Introduction}\\

\bigskip

Nonlinear systems yield a rich gamut of dynamical behaviors that range from fixed points and limit cycles of varying periodicities, to chaotic attractors. In this work we will exploit the presence of dynamical attractors localized in different regions of phase space, and the possibility of hopping between such attractors, to obtain logic operations. 

Consider a general nonlinear system of the form:
\begin{eqnarray}
\label{mlc}
\dot{x}&=&y - g(x) , \\ \nonumber
\dot{y}&=& - a y - x + b + I + f(t) ,
\end{eqnarray}
where $f(t)$ is a periodic forcing signal, $g(x)$ is a nonlinear function, $b$ is a constant bias and $I$ is an input signal. Specifically, consider a simple easily implementable piecewise linear form for
\begin{equation}
  \label{gx}
g(x) = c_1 x + \frac{1}{2} (c_2-c_1) (|(x+1)|-|(x-1)|)
\end{equation}
and $f(t) = A \sin (\omega t)$, where $\omega$ is the frequency and $A$ is the amplitude of the periodic forcing. These dimensionless coupled first order differential equations underlie the readily implementable MLC circuit \cite{mlc}.

\begin{figure}[ht]
\begin{center}
  \includegraphics[width=1\linewidth]{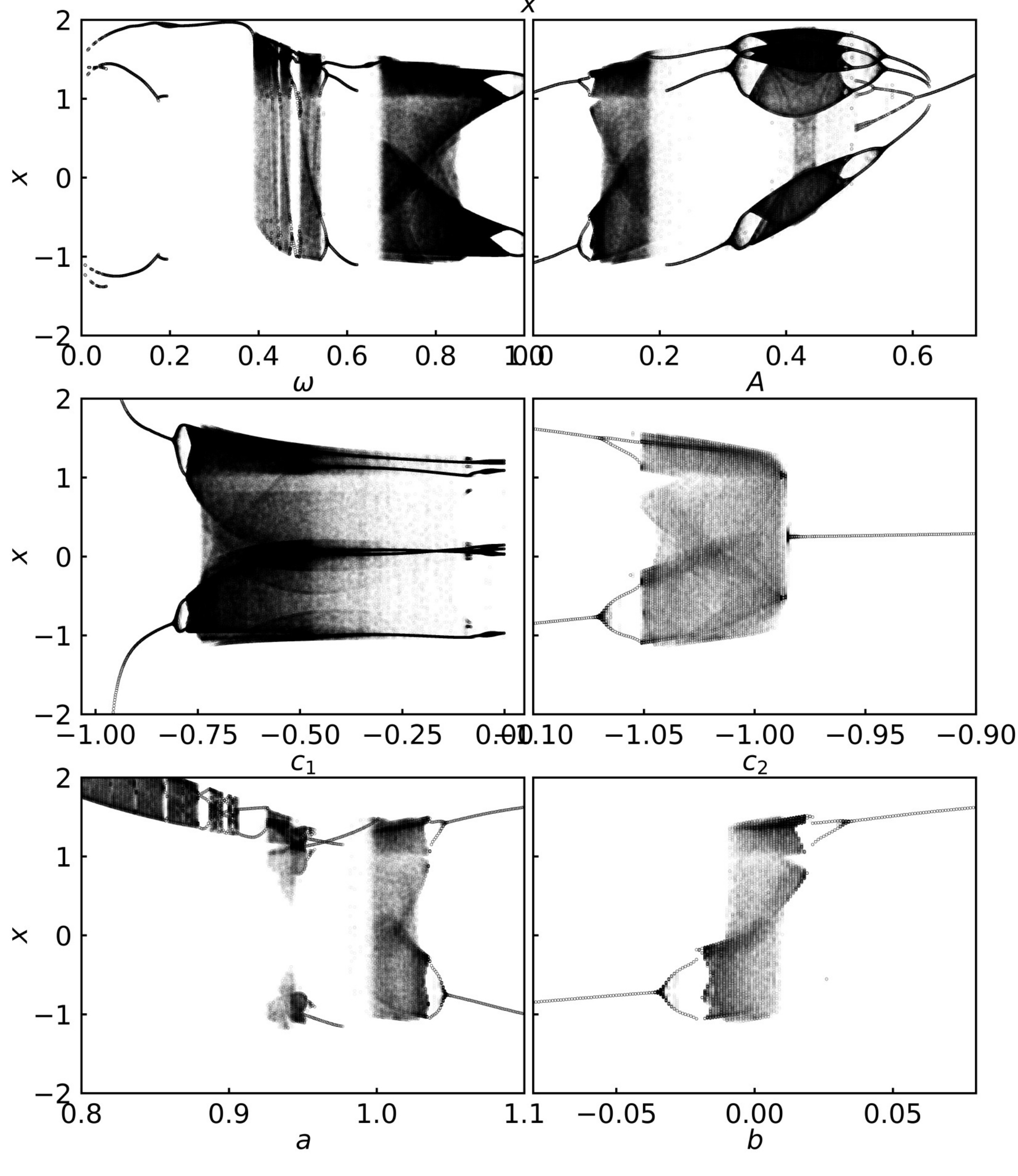}
\end{center}
\caption{
Bifurcation diagrams displaying the distinct dynamical attractors obtained with respect to the different parameters in the system (cf.  Eqns.~\ref{mlc}-\ref{gx}): frequency $\omega$, amplitude $A$, $c_1$, $c_2$, $a$ and bias parameter $b$.
While one of the parameters is being varied, the other parameters (whichever are relevant) are fixed at: $a=1.015$, $c1=-0.55$, $c2=-1.02$, $\omega=0.74$, $A=0.11$, $b=0$ in Eqn.~\ref{mlc}. 
}
\label{bif_diag}
\end{figure}

The bifurcation diagrams of the system
  with respect to all the different parameters are shown in
  Fig.~\ref{bif_diag}, depicting the richness of behaviors which may
  be exploited for implementing different logic
  operations. Specifically we seek attractors in parameter
  space that occupy clearly distinct regions. The most suitable
  parameter that offers this feature, as well as the simplest one to
  manipulate, is the bias parameter $b$. So in this work we will use the
  patterns evident in the bifurcation diagram of the system, with
  respect to bias $b$, to design logic gates.

In order to conceive of a mapping of the dynamics to logic operations,
we need to specify the inputs-to-output correspondence. We first focus
on the encoding of logic inputs. In general, $N$ logic inputs are
encoded by $N$ square waves which constitute the input signal $I$ in
Eqn.~\ref{mlc}.  In particular, for two logic inputs, the input signal
$I$ is the sum $I_1 + I_2$, with $I_1$ and $I_2$ encoding the two
logic inputs.  Since the logic inputs can be either $0$ or $1$, they
can combine to give four logic input sets $(I_1, I_2)$: $(0,0)$,
$(0,1)$, $(1,0)$ and $(1,1)$, with the input sets $(0,1)$ and $(1,0)$
giving rise to the same $I$. This implies that the four input
conditions $(I_1, I_2)$ reduce to three distinct values of $I$. Hence,
the input signal $I$, generated by adding two independent uncorrelated
input signals, is a $3$-level aperiodic waveform. In this work the
input signal $I$ will be considered to be of {\em very low amplitude},
compared to the typical size of the chaotic attractor. The central
idea here rests on the capability of the nonlinear system to yield a
large response, such as a very different dynamical attractor, in
response to a very small input signal.

Now, this nonlinear system is capable of exhibiting attractors that are bounded in different regions of phase space. For instance, it can give rise to attractors where the value of the $x$ (or $y$) variable is entirely positive, as well as attractors whose $x$ (or $y$) values are entirely negative, under variation of the small input signal $I$. Dynamically, these attractors may be fixed points, periodic cycles or even chaotic attractors. So as the value of $I$ switches, i.e. the input set switches, we observe that the attractors can jump from a certain sector of phase space to a very different sector. This is the feature which we will exploit to implement a robust input-output correspondence in this system \cite{ss1,ss3,ss4,ss5,ss6,ss7,ss8,ss9,pra}. 

So the dynamical attractor of the system will yield the logic output. For instance, if $x(t)$ (or $y(t)$) is greater than $x_{thresh}$ or $y_{thresh})$ respectively, it is mapped to logic output $1$, and if $x(t)$ (or $y(t)$) is lower than $x_{thresh}$ or $y_{thresh})$ respectively, it is mapped to logic output $0$. The thresholds for output determination $x_{thresh}$ and $y_{thresh}$ can be suitably chosen, and are typically close to zero. As mentioned earlier, specifically, we can have $x_{thresh}=y_{thresh}=0$, namely we can consider the output to be a logical $1$ if the inputs yield a positive attractor, and the output to be a logical $0$ if it is a negative attractor, i.e. if $x ({\rm or} \ y) < 0$, Logic Output is $0$ and if $x ({\rm or} \ y) > 0$, Logic Output is $1$.

\begin{figure}[ht]
\begin{center}
\includegraphics[width=1\linewidth]{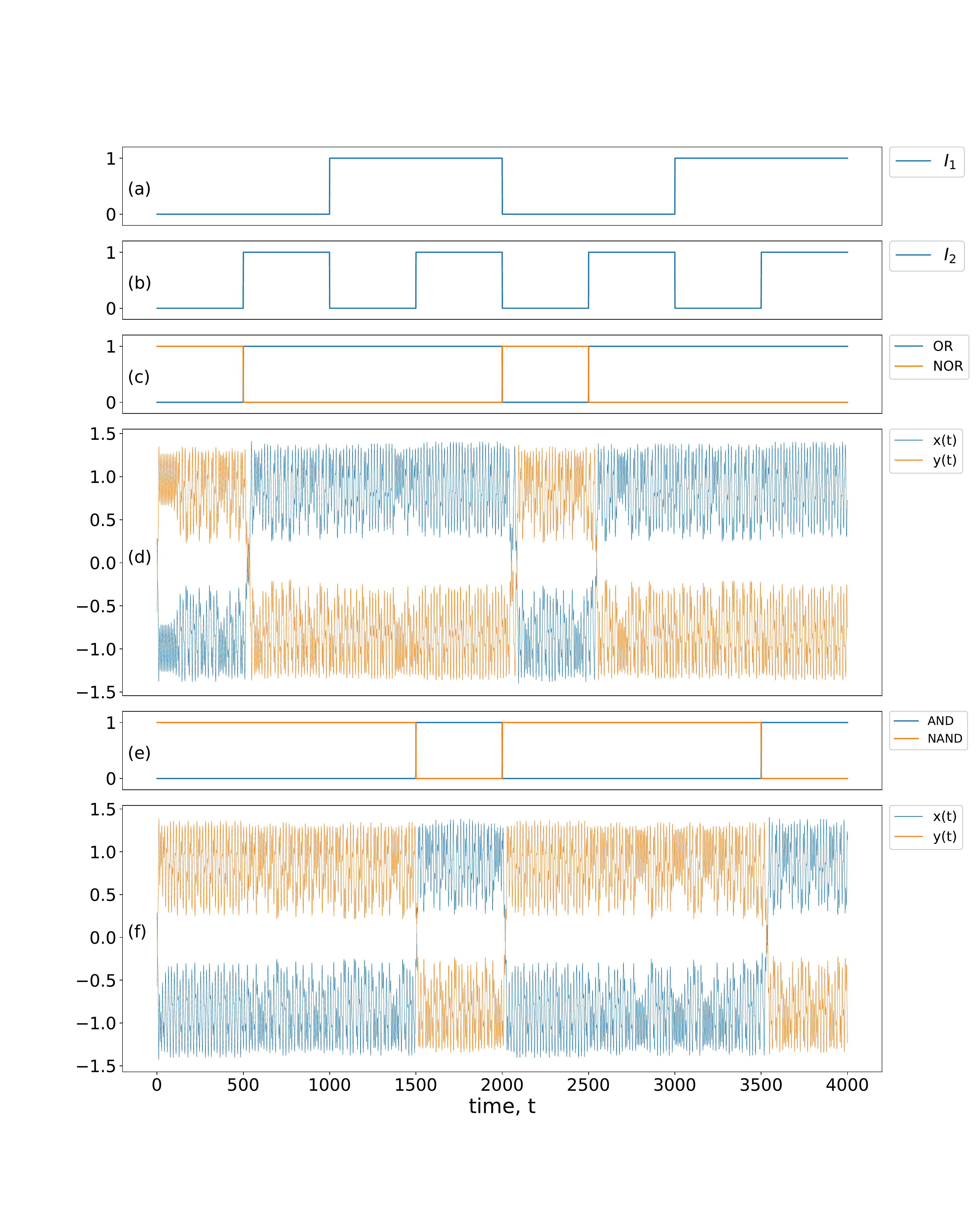}
\end{center}
\caption{Logic output from attractor hopping: Panels (a) and (b) show a stream of inputs $I_1$ and $I_2$.  Panel (c) shows the expected OR/NOR logic output and (e) shows the AND/NAND logic output corresponding to logic input set ($I_1$, $I_2$).  Panels (d) and (f) show outputs $x(t)$ and $y(t)$ of the nonlinear system given by Eqn.~\ref{mlc} with $a = 1.015$, $c_1=-0.55$, $c_2 = -1.02$, $A = 0.11$, $\omega = 0.74$. The input signals take value $-0.002$ when logic input is $0$ and value $0.002$ when logic input is $1$, and logic output is $1$ when  $x(t)$ ( or $y(t)$ ) $> 0$, and logic output is $0$ when $x(t)$ ( or $y(t)$ ) $< 0$. In (d), bias $b = -0.002$, and clearly the $x$ variable yields a consistent logical OR output, while the $y$ variable yields a consistent NOR logic output. In (f), bias $b = +0.002$, and clearly the $x$ variable yields a consistent logical AND output, while the $y$ variable yields the complementary NAND logic gate.
}
\label{timing_mlc_sim}
\end{figure}

\begin{figure}[ht]
\begin{center}
    \includegraphics[width=1\linewidth]{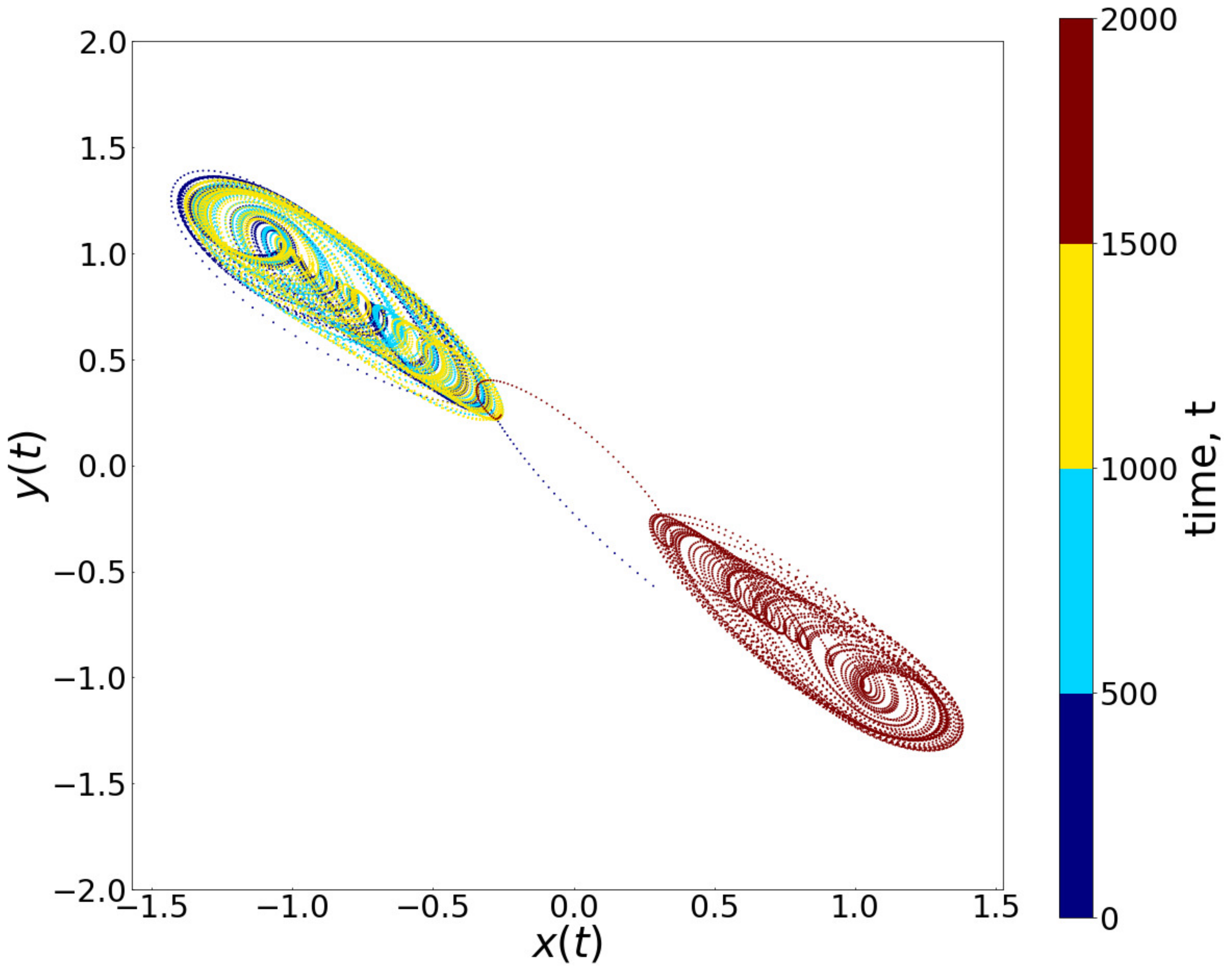}
\end{center}
\caption{Phase portraits of the dynamical attractors arising from the  different input sets in Fig.~\ref{timing_mlc_sim} (with  $a = 1.015$, $c_1=-0.55$, $c_2 = -1.02$, $A = 0.11$, $\omega = 0.74$ in Eqn.~\ref{mlc}). Here the time evolution under different input sets is depicted in different colors. It is clearly evident that the trajectory switches between chaotic attractors in two distinct quadrants as the logic input sets change, yielding appropriate output states.}
\label{attractors}
\end{figure}

We will now demonstrate here that a given set of inputs ($I_1,I_2$)
yields an output in accordance with the truth tables of the basic
logic operations shown in Table~1. Crucially, the different outputs
will arise from the {\em chaotic attractor hopping induced by the
  input stream}. We present explicit examples of this phenomenon, from
numerical simulations, in Figs.~1-2. These figures show illustrative
cases of positive and negative chaotic attractors yielding Logic
Output $1$ and $0$ respectively, under a stream of external input
signals. So as the system receives different inputs it switches
between these qualitatively different dynamical attractors, yielding
the appropriate output. Fig.~\ref{attractors} specifically shows the
two one-band/single-scroll chaotic attractors that occupy distinct
regions of phase space, characterizing the two outputs. So, as the
system hops between these chaotic attractors, the output toggles
between $0$ and $1$. Significantly, the very low-amplitude input
signals yield highly amplified outputs. For instance, in our
representative example, the input signal $I=0.002$ results in
dynamical attractors that differ on average by $\sim 2$,
i.e. approximately two orders of magnitude larger. Namely, the {\em
  input signal (which is of the order of $10^{-3}$) is very small,
  vis-a-vis the size of chaotic attractor (which is of the order of
  $1$),} implying that a small change in the system yields a
significantly different dynamical outcome. Also note
  that the response time of the system is very short, with the system
  taking of the order of microseconds on average to switch between the
  desired states, leading to low latency.

Further, under a different bias parameter $b$ in
Fig.~\ref{timing_mlc_sim}(f), we also obtain a consistent OR logic
operation, again by switching between chaotic attractors confined in
distinct quadrants of phase space. So the system has the capablity of
implementing different logic operations {\em flexibly} through a
simple change of bias parameter, leading to potentially reconfigurable
logic gates.

Additionally, the $x$ and $y$ variables yield {\em complementary}
logic outputs in parallel. So in the specific examples presented in
Figs.~\ref{timing_mlc_sim}d and f, variable $x$ yields a consistent
AND/OR gate response, while variable $y$ yields a consistent NAND/NOR
gate response. Thus this two-dimensional system allows us to {\em
  implement operations in parallel} by simultaneously yielding the two
complementary logic outputs.

\begin{table}
~\hfill\begin{tabular}{|c|*{16}{c|}}\hline
Input Set ($I_1$,$I_2$)&OR&AND&NOR&NAND\\ \hline
\hline
(0,0) &0&0&1&1\\ 
(0,1)/(1,0) &1&0&0&1\\ 
(1,1) &1&1&0&0\\ \hline
\end{tabular}\hfill~
\caption{\small Relationship between the two inputs and the output of the fundamental OR, AND, NOR and NAND operations. Note that the four distinct possible input sets $(0,0)$, $(0,1)$, $(1,0)$ and $(1,1)$ reduce to three conditions as $(0,1)$ and $(1,0)$ are symmetric. Note that {\em any} logical circuit can be constructed by combining the fundamental NOR (or the NAND) gates \cite{mano}. }
\end{table}

\bigskip
\bigskip
	

\noindent
{\bf Quantifying the reliability of obtaining a logic output}\\

\bigskip

We can quantify the consistency (or reliability) of
  obtaining a given logic output by calculating the probability,
  denoted by $P (logic)$, of obtaining the desired logic output for
  different input sets. $P (logic)$ is estimated from numerical
  simulations by calculating the ratio of the number of successful
  runs to the total number of runs. Each run consists of a permutation
  of the inputs sets $(0,0)$, $(0,1)/(1,0)$, $(1, 1)$ fed sequentially
  to the system. If the logic output obtained from the system is the
  desired output for {\em all} input sets in the run, it is considered
  a ``success''. Even if one input set returns a wrong output, the
  full run is considered a ``failure''. So this quantity offers a very
  stringent measure of reliability.  When $P (logic)$ is $1$, the
  logic operation is very reliable, as the system yields the correct
  output in response to all the input sets ($I_1$,$I_2$) provided to
  it. Specifically in the numercial results presented here we sample
  $10^3$ runs. 

\begin{figure}[ht]
\begin{center}
\includegraphics[width=1\linewidth]{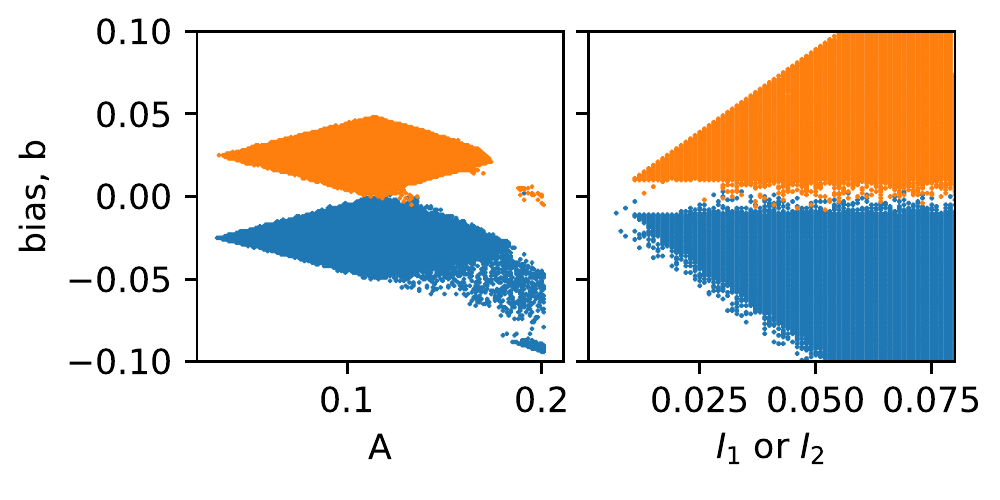}
\end{center}
\caption{Regions in the parameter space of bias $b$ ($y$-axis) and forcing amplitude $A$ ($x$-axis, left), input $I_1/I_2$ ($x$-axis, right), and where the probability of obtaining NAND (orange) and NOR (blue) logic is $1$. Here $a=1.015$, $c_1=-0.55$, $c_2=-1.02$, $\omega=0.75$ and the inputs $I_1/I_2$ take values $-0.025/0.025$ for logic input $0/1$ in (a) and $A=0.14$ in (b). Here the dynamical attractors may be limit cycles or chaotic attractors.}
\label{PNOR_A_b}
\end{figure}

It is evident from Fig.~\ref{PNOR_A_b}, which shows $P (logic)$
obtained from numerical simulations, that the fundamental logic
operations NOR and NAND are realized consistently in an optimal band
of moderate forcing amplitude $A$ and bias $b$. The logic response is
100\% accurate in this reasonably wide window. That is, in a broad
region of parameter space the system yields outputs, in response to
external inputs, in complete accordance with the fundamental logic
functionalities shown in the truth tables (cf. Table~1). Also,
importantly, a simple switch in bias $b$ changes the logic gate from
NOR/OR to NAND/AND. This implies that the system can operate flexibly
as a NOR/OR logic gate or a NAND/AND logic gate, with the small bias
parameter having the capacity to morph the system to operate as
different logic gates.

\bigskip
\bigskip

\noindent
{\bf Experimental Verification}\\

\bigskip

\begin{figure}[htb]
\begin{center}
\includegraphics[width=0.65\linewidth,angle=270]{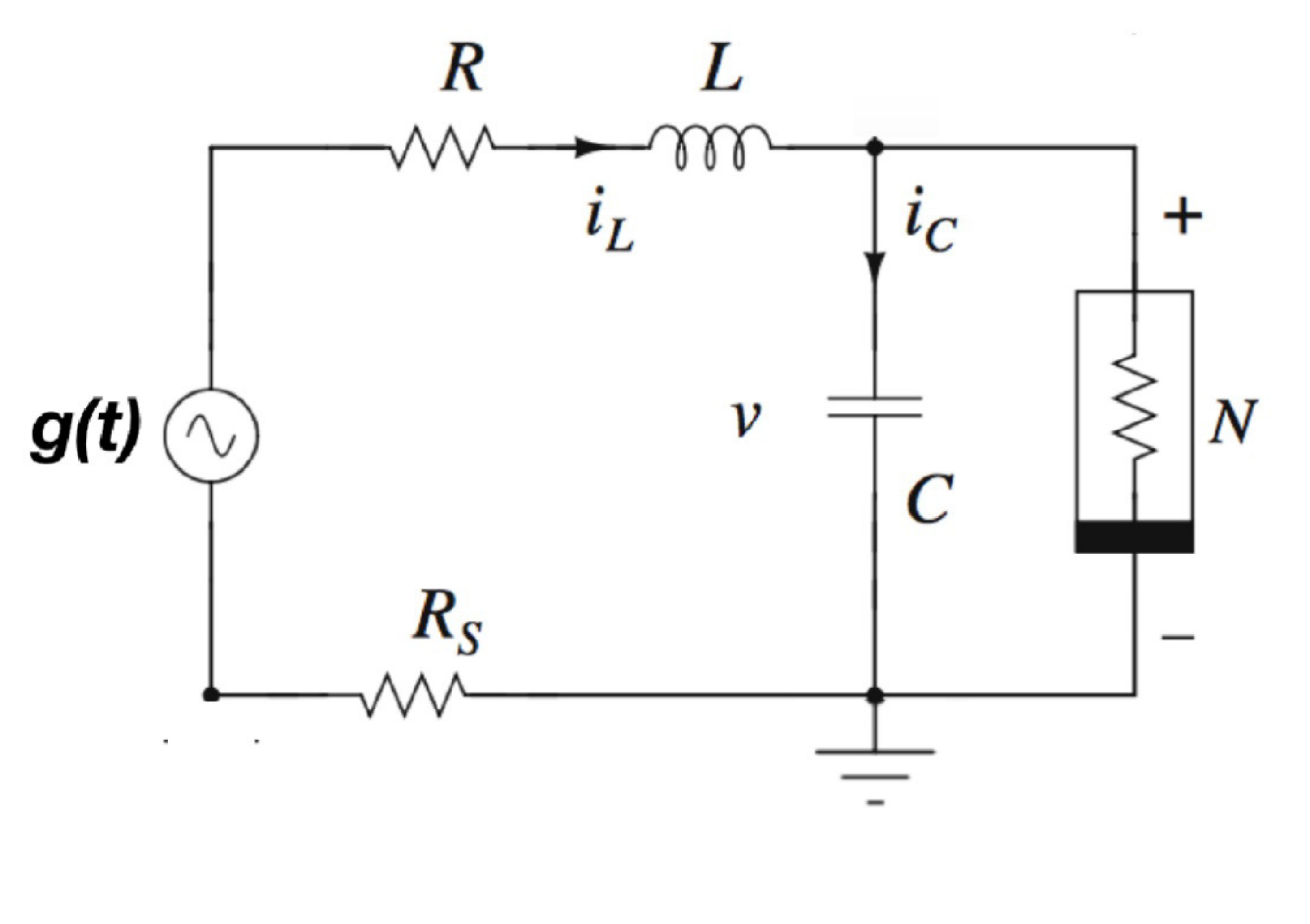}
\end{center}
\caption{Schematic of a simple electronic circuit, known as the MLC
  circuit \cite{mlc}, implementing the rescaled dynamical equations
  given in Eqn.~\ref{mlc}. In the circuit, the voltage across the
  capacitor $C$, the current through the inductor $L$ and the external
  forcing signal $g(t)$ correspond to $x$, $y$ and $b+I+f(t)$ in
  Eqn.~\ref{mlc} respectively. The nonlinear element $N$ is the Chua’s
  diode implemented as in \cite{mlc}, with rescaled piecewise-linear characteristic curve
  $g(x) = c_1 x + \frac{1}{2} (c_2-c_1) (|(x+1)|-|(x-1)|)$.}
\label{fig5}
\end{figure}

\begin{figure}[htb]
  \begin{center}
      \includegraphics[width=0.65\linewidth]{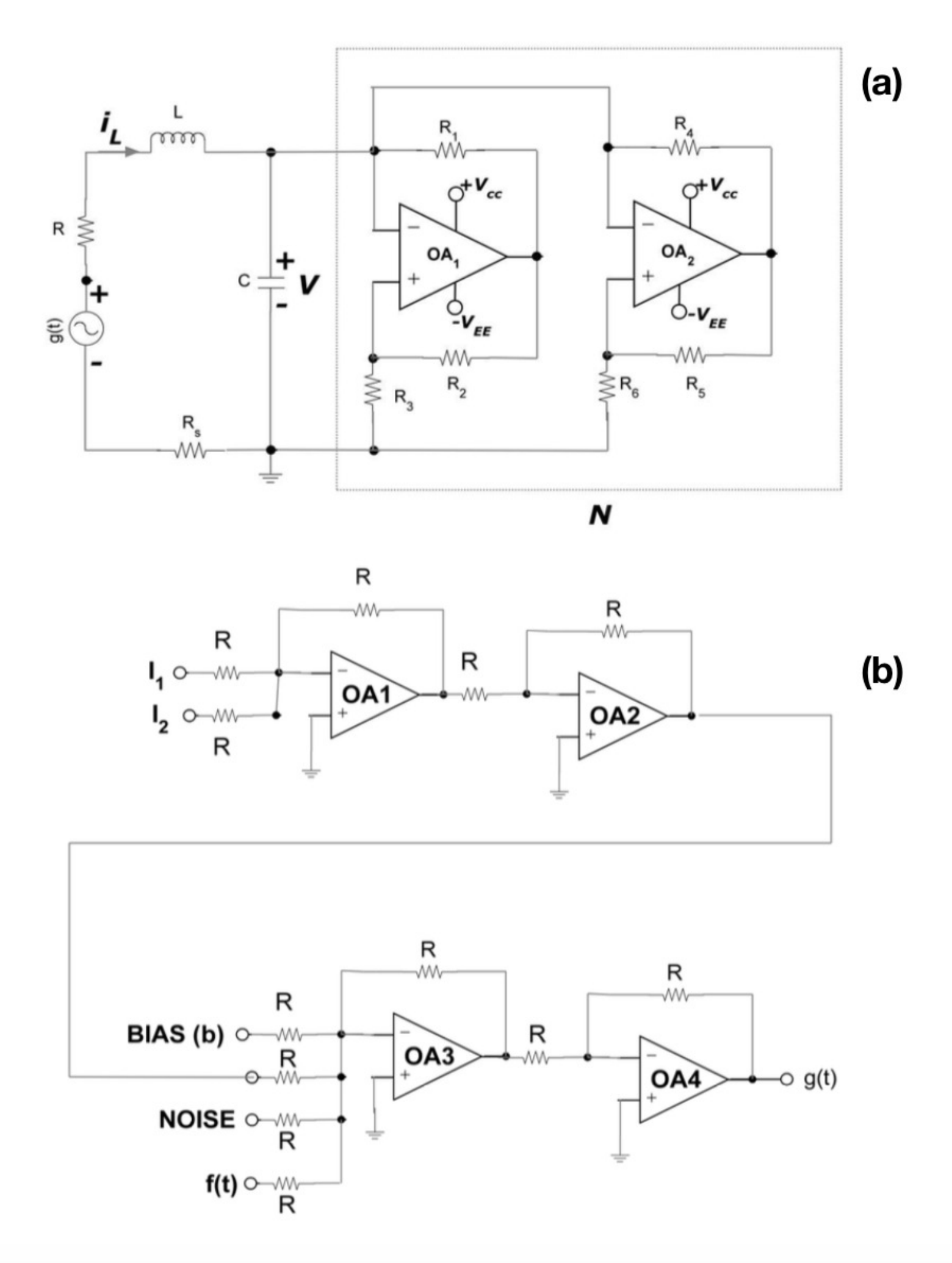}
\end{center}
  \caption{(a): Realization of MLC circuit using two
      op-amps (AD712, TL082, AD844, or equivalent) and six linear
      resistors to implement the Chua’s diode ($N$).  The resistors $R
      = 1340 \ \Omega$, $R_1 = R_2 = 22 \ k\Omega$, $R_3 = 3.3 \ k\Omega$,
      $R_4 = R_5 = 220 \ \Omega$, $R_6 = 2.2 \ k\Omega$ and $R_s = 20
      \ \Omega$. The capacitor $C = 10 \ nF$ and the inductor $L = 18 \ mH$
      (TOKO type 10RB or equivalent). Here $g(t)$ is the input driving
      signal, the capacitor voltage is $V(t)$ and the inductor current
      is $i_L$. The current $i_L$ is sensed through the resistor
      $R_s$ to give the voltage $V_L$ \cite{mlc}. (b) Circuit for
      generating the driving signal $g(t)$. Here op-amps OA1 - OA4 are
      realized with AD712. All the resistors are fixed as $R = 10
      \ k\Omega$. The power-supply to op-amps and the bias voltage
      ($b$) are drawn from Agilent or Keysight E3631A DC Power
      Supply. The sinusoidal signal $f(t)$ and the noise signal are
      drawn from Agilent or Keysight 33522A, Function/Arbitrary
      Waveform Generator.}
\label{fig5ab}
\end{figure}

Now we will verify this concept in electronic circuit analogs of the
nonlinear system described by Eqn.~\ref{mlc}, and ascertain its
robustness in experiments. The schematic of the
  circuit realization of the simple non-autonomous MLC circuit is
  shown in Fig.~\ref{fig5}. It contains a capacitor, an inductor, a
  linear resistor, an external periodic forcing $g(t)$ and only one
  nonlinear element, namely, the Chua’s diode ($N$) \cite{mlc}. The
  complete circuit implementation of MLC circuit is depicted in
  Fig.~\ref{fig5ab}(a). To measure the inductor current $i_L$ in our
  experiments, we insert a current sensing resistor $R_s$ as shown in
  Fig.~\ref{fig5ab}a to give the voltage $V_L$ \cite{mlc}. By applying
  Kirchhoff's laws to this circuit, the governing equations for the
  voltage ($V$) across the capacitor $C$ and the current $i_L$ through
  the inductor $L$ are represented by the rescaled Eqns.~1
  \cite{mlc}. Two op-amps (AD712, TL082, AD844, or equivalent) and six
  linear resistors are used to implement the Chua's diode ($N$). The
  parameters of the circuit elements are fixed as resistors $R = 1340
  \ \Omega$ , $R_1 = R_2 = 22 \ k\Omega$, $R_3 = 3.3 \ k\Omega$, $R_4
  = R_5 = 220 \ \Omega$, $R_6 = 2.2 \ k\Omega$ and $R_s = 20
  \ \Omega$. The capacitor $C = 10 \ nF$ and the inductor $L = 18
  \ mH$ (TOKO type 10RB or equivalent). The frequency of the external
  sinusoidal force $f(t)$ as in Fig.~\ref{fig5ab}(b) is fixed at $8890
  \ Hz$.  The circuit of Fig.~\ref{fig5ab}(b) is used to generate the
  driving signal $g(t)$ for the circuit of Fig.~\ref{fig5ab}(a). In
  the circuit of Fig.~\ref{fig5ab}(b), $g(t)$ is basically generated
  by a set of op-amp summing amplifiers by adding the logic input
  signals $I_1$ and $I_2$, external bias voltage ($b$), external noise
  signal and the sinusoidal signal $f(t)$. All the op-amps are
  employed with AD712 (or TL082 or AD844 or equivalent). The voltage
  supply for all the op-amps are fixed at $\pm 9 \ V$. All the
  resistors are fixed as $R = 10 \ k\Omega$. 

\begin{figure}[ht]
  \begin{center}
\includegraphics[width=1\linewidth,height=0.75\linewidth]{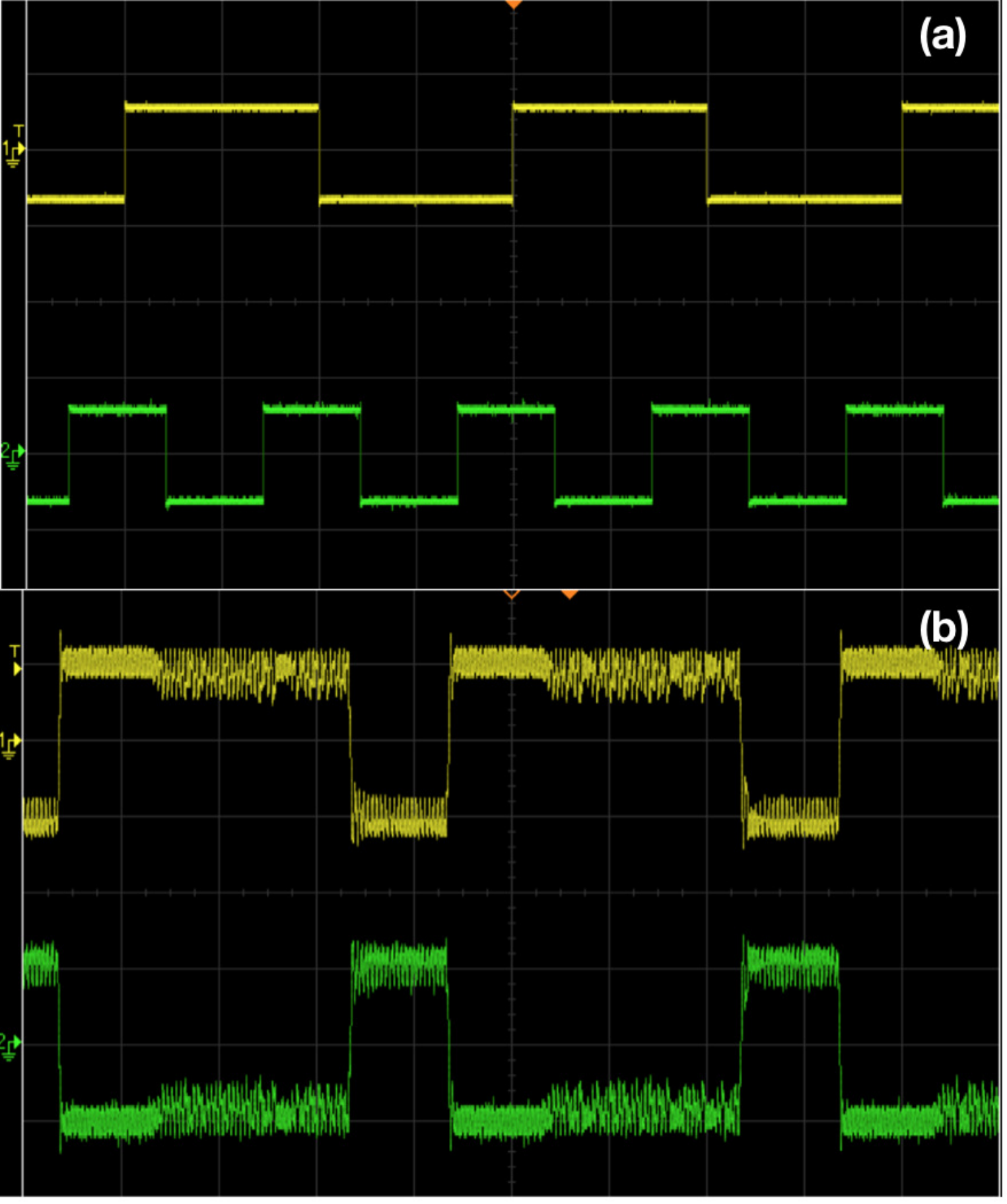}
\end{center}
\caption{Realization of OR/NOR logic gates through chaotic attractor hopping
in electronic circuit experiments: Panel (a) shows stream of inputs $I_1$
and $I_2$ (which take value $-10 \ mV$ when logic input is $0$ and value $10 \ mV$ when logic input
is $1$). Panel (b) shows the voltage $V(t)$ (yellow) clearly indicating a logical OR output
(with $V(t) > 0$ being logic output $1$, and $V(t) < 0$ being logic output $0$). The output voltage
$V_L$ (green) yields the complementary NOR logic gate response. The amplitude $A$ of the
sinusoidal signal is $150 \ mV$ and frequency is $8890 \ Hz$. The bias voltage $b$, is fixed as
$10 \ mV$. For panel (a), the scale in the traces are: $20 \ mV$/Div ($Y$-axis) and $5 \ mS$/Div ($X$-
axis). For panel (b), the scale in the traces are: $100 \ mV$/Div ($Y$-axis) and $5 \ mS$/Div ($X$-
axis). The oscilloscope used is Agilent or Keysight DSOX2012A.}
\label{expt_AND_chaotic}
\end{figure}

\begin{figure}[ht]
\begin{center}
  \includegraphics[width=1\linewidth,height=0.75\linewidth]{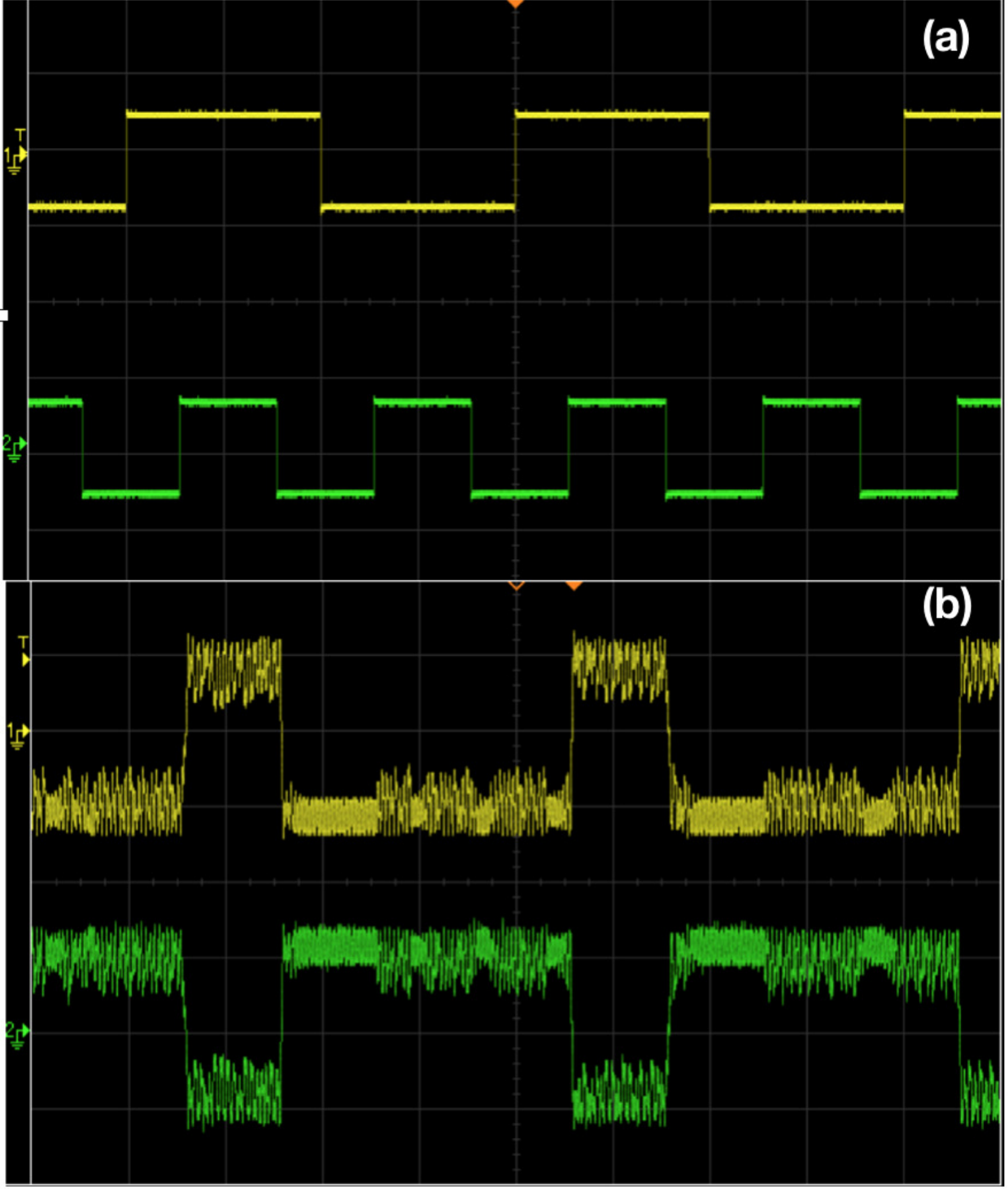}
\end{center}
\caption{Realization of AND/NAND logic gates through chaotic attractor hopping
in electronic circuit experiments: Panel (a) shows stream of inputs $I_1$
and $I_2$ (which take value $-10 \ mV$ when logic input is $0$ and value $10 \ mV$ when logic input
is $1$). Panel (b) shows the voltage $V(t)$ (yellow) clearly indicating a logical AND output
(with $V(t) > 0$ being logic output $1$, and $V(t) < 0$ being logic output $0$). The output voltage
$V_L$ (green) yields the complementary NAND logic gate response. The amplitude $A$ of the
sinusoidal signal is $150 mV$ and frequency is $8890 \ Hz$. The bias voltage $b$, is fixed as
$-10 \ mV$. For panel (a), the scale in the traces are: $20 \ mV$/Div ($Y$-axis) and $5 \ mS$/Div ($X$-
axis). For panel (b), the scale in the traces are: $100 \ mV$/Div ($Y$-axis) and $5 \ mS$/Div ($X$-
axis)}
\label{A_dependence}
\end{figure}

As indicated by the numerical simulations in Fig.~\ref{PNOR_A_b}, the
amplitude of the forcing has to be in an optimal moderate range to
obtain logic operations. Figs.~\ref{expt_AND_chaotic}-\ref{A_dependence} verify
this behavior in electronic circuits. When the forcing amplitude is
too small, the system tends to get stuck in some region of phase space
and is unable to hop to the appropriate attractor. On the other hand,
too large forcing amplitude results in the system hopping randomly
from one sector of phase space to another, due to underlying double
scroll attractors. Clearly, the intermediate forcing amplitude yields
consistent logic operation, with appropriate attractor hopping induced
only by changes in the input signal.

\bigskip
\bigskip
\newpage

\noindent
{\bf Influence of Noise: Generalized Logical Stochastic Resonance}\\

\bigskip

Lastly, we will investigate the effect of noise on the logic responses
of the system \cite{sr,sr1}. The first issue is to ascertain the
robustness of the logic response with respect to ambient noise,
i.e. we will check if noise degrades performance, or not. Secondly, we
would like to investigate if there are some regions of dynamical
behavior where noise aids the reliability of obtaining the correct
logic output. In earlier studies it has been shown that a bistable
system supporting two fixed point states, driven by a stream of
sub-threshold input signals, yields enhanced probability of correct
logic responses, in a window of optimal noise. This phenomena has been
called {\em Logical Stochastic Resonance} (LSR)
\cite{lsr1,lsr2,lsr3,lsr4,lsr6,lsr7}, and it has been realized in
systems ranging from nanomechanical oscillators \cite{nano},
coulomb-coupled quantum dots \cite{qdots} and optical systems
\cite{lsr_optics,lsr_optics1} to chemical systems \cite{lsr_chem} and
synthetic genetic networks
\cite{sgn1,sgn11,sgn2,sgn3,sgn4,sgn5,sgn6}. Extensions
  of the idea to include the effect of periodic forcing was
  demonstrated in \cite{lsr8}, where the width of the optimal noise
  window was shown to increase by utilizing periodic forcing,
  i.e. noise in conjunction with a periodic drive yielded consistent
  logic outputs for all noise strengths below a certain
  threshold. Further, the LSR concept has been demonstrated in coupled
  systems and higher-dimensional systems, with multiple steady states
  \cite{sgn1,lsr5,lsr9}. Specifically, two complementary gate
  operations were achieved simultaneously in a two-dimensional model of
  a gene network \cite{sgn1}, indicating the flexible parallel
  processing potential of a biological system. In another direction
  for two coupled systems it was demostrated that, even when the
  individual systems receive only one logic input each, due to the
  interplay of coupling, nonlinearity and noise, they cooperatively
  respond to give a logic output that is a function of both inputs
  \cite{lsr9}. Further, the idea was extended to multi-stable
  dynamical systems with more than two stable fixed points, allowing
  one to obtain XOR logic, in addition to the AND (NAND) and OR (NOR)
  logic observed earlier \cite{lsr5}.

Now in all its variations (some of which are detailed
  above), the concept of Logical Stochastic Resonace has so far been
  restricted only to {\em steady states}. In this work we explore the
  scope of the idea of LSR for the case of more complex attractors
  such as periodic cycles, or even {\em chaotic} attractors. Our
basic question is then as follows: does the idea of Logical Stochastic
Resonance extend beyond fixed point states, to more complex dynamical
attractors? If it does indeed hold for more complex dynamics, we will
have attained a {\em generalized} concept of Logical Stochastic
Resonance.

\begin{figure}[ht]
\begin{center}
  \includegraphics[width=1\linewidth,height=1\linewidth]{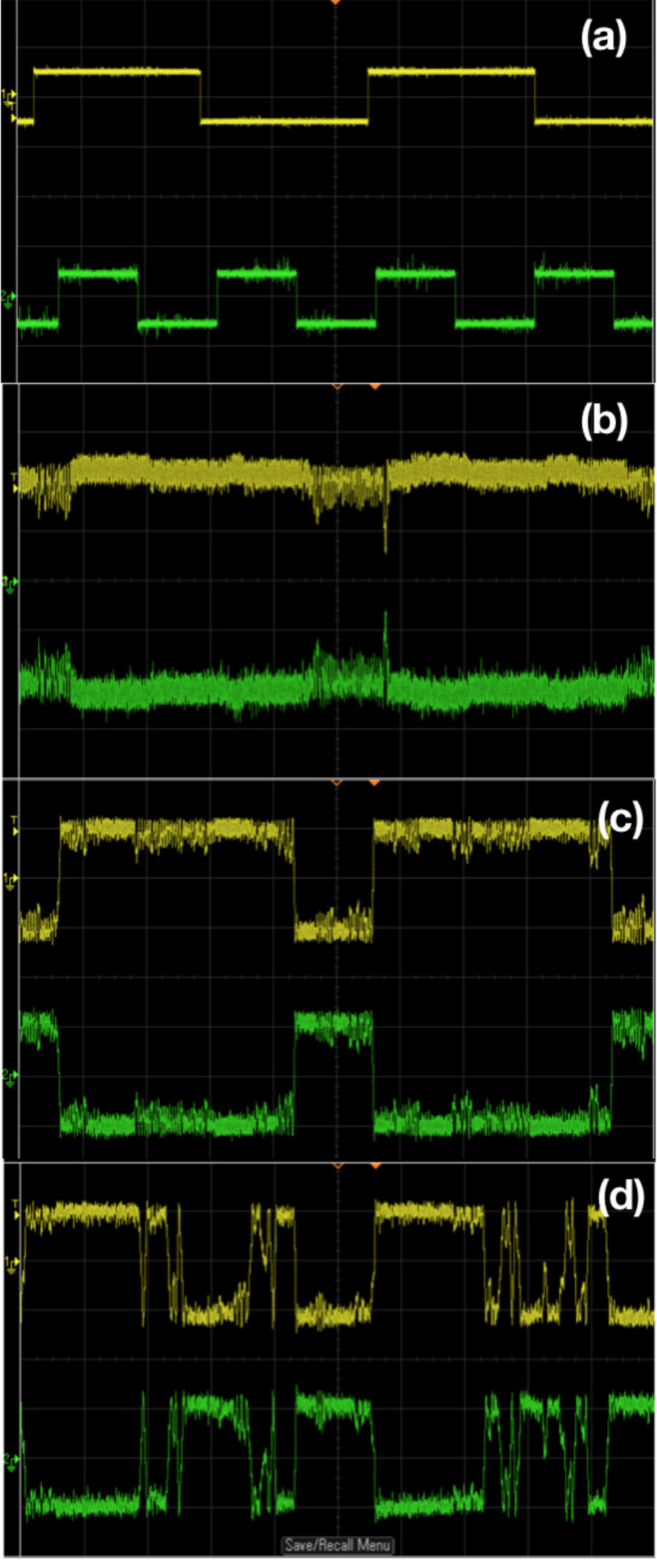}
\end{center}
\caption{Realization of the OR/NOR logic gate at intermediate
    noise strengths in electronic circuit experiments: Panel (a) shows
    stream of inputs $I_1$ and $I_2$ (which take value $-10 \ mV$ when
    logic input is $0$ and value $10 \ mV$ when logic input is
    $1$). Panels (b) to (d) show the output voltage $V(t)$ (yellow)
    and $V_L (t)$ (green) for different noise strengths $D$: (i) $100
    \ mV$, (ii) $1.0 \ V$ and (iii) $1.5 \ V$. Clearly panel (c)
    depicts logical OR output (with $V(t) > 0$ being logic output $1$,
    and $V(t) < 0$ being logic output $0$).  The output voltage $V_L
    (t)$ (green) yields the complementary NOR logic gate response. The
    amplitude $A$ of the sinusoidal signal is $100 \ mV$ and frequency
    is $8890 \ Hz$. The bias voltage $b$, is fixed as $10 \ mV$. For
    panel (a), the scale in the traces are: $20 \ mV$/Div ($Y$-axis)
    and $5 \ mS$/Div ($X$-axis). For panel (b-d), the scale in the
    traces are: $100 \ mV$/Div ($Y$-axis) and $5 \ mS$/Div
    ($X$-axis). For panel (a), the scale in the traces are: $20
    \ mV$/Div ($Y$-axis) and $5 \ mS$/Div ($X$-axis). For panel (b-d),
    the scale in the traces are: $100 \ mV$/Div ($Y$-axis) and $5
    \ mS$/Div ($X$-axis).}
\label{noise}
\end{figure}

Fig.~\ref{noise} shows representative experimental results of this, for the
system in Eqn.~\ref{mlc} under additive zero-mean Gaussian noise, given as:
\begin{eqnarray}
\label{mlc_noise}
\dot{x}&=&y - g(x) ,\\ \nonumber
\dot{y}&=& - a y - x + b + I + f(t) + D \eta (t) ,
\end{eqnarray}
where $\eta (t)$ is a zero-mean Gaussian noise with variance $1$, and
parameter $D$ gives the noise strength. In the circuit implementation displayed in Fig.~\ref{fig5}, $g(t)$ now corresponds to $b+ I + f(t) + D \eta(t)$.

\begin{figure}[ht]
\begin{center}
	\includegraphics[width=1\linewidth]{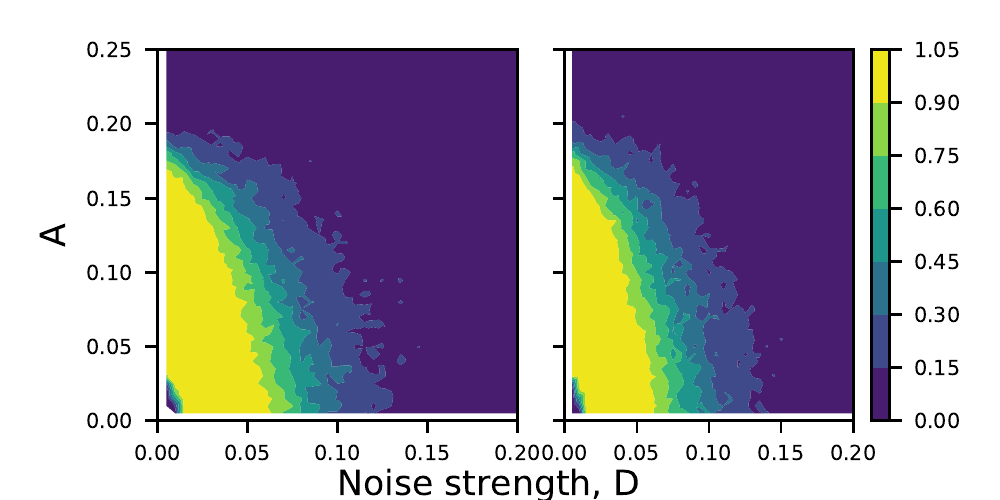}
\end{center}
\caption{Probability of obtaining NOR logic, obtained from numerical
  simulations (with $b=0.025$, on the left) and NAND logic (with
  $b=-0.025$, on the right) in the parameter space of forcing
  amplitude $A$ ($y$-axis) and noise strength ($x$-axis) in
  Eqn.~\ref{mlc_noise}, with $\omega=0.75$ and inputs $I_1/I_2$ take
  value $-0.025$ when logic input is $0$ and value $0.025$ when logic
  input is $1$.}
\label{PNOR}
\end{figure}

Now, the forcing amplitude in the case illustrated is too small to
yield appropriate attractor hopping that may be mapped to the output
desired for logic operations, for subthreshold input
signals. Naturally, when noise strength is too large, the system jumps
randomly between attractors, and thus the system cannot yield any
reliable output. When noise is zero or too small the system is
essentially stuck in one dynamical attractor. However, remarkably,
robust logic operations are realized when there is some noise in the
system. So in the presence of moderate noise the system jumps from
attractor to attractor in response to inputs consistently. Since these
attractors are more complex than fixed points considered in earlier
studies, these results offer a significant generalization of the
concept of Logical Stochastic Resonance. The quantification of the
reliability of obtaining a logic output through Logical Stochastic
Resonance is depicted in Fig.~\ref{PNOR}. It is clear that in
relatively wide windows of moderate noise, the system yields logic
operations with near certain probability i.e., $P(logic)\sim 1$.  {\em
  Remarkably, note that the amplitude of the logic input signal is
  very low here, and may even be smaller than the noise strength.}
For instance, in the particular illustrative example
  displayed in Fig.~\ref{noise}, the input signal ($I = 10 mV$) is
  $100$ times smaller than the typical experimental noise strength in
  the optimal window of noise ($\sim 1 V$).

\begin{figure}[ht]
\begin{center}
	\includegraphics[width=0.8\linewidth]{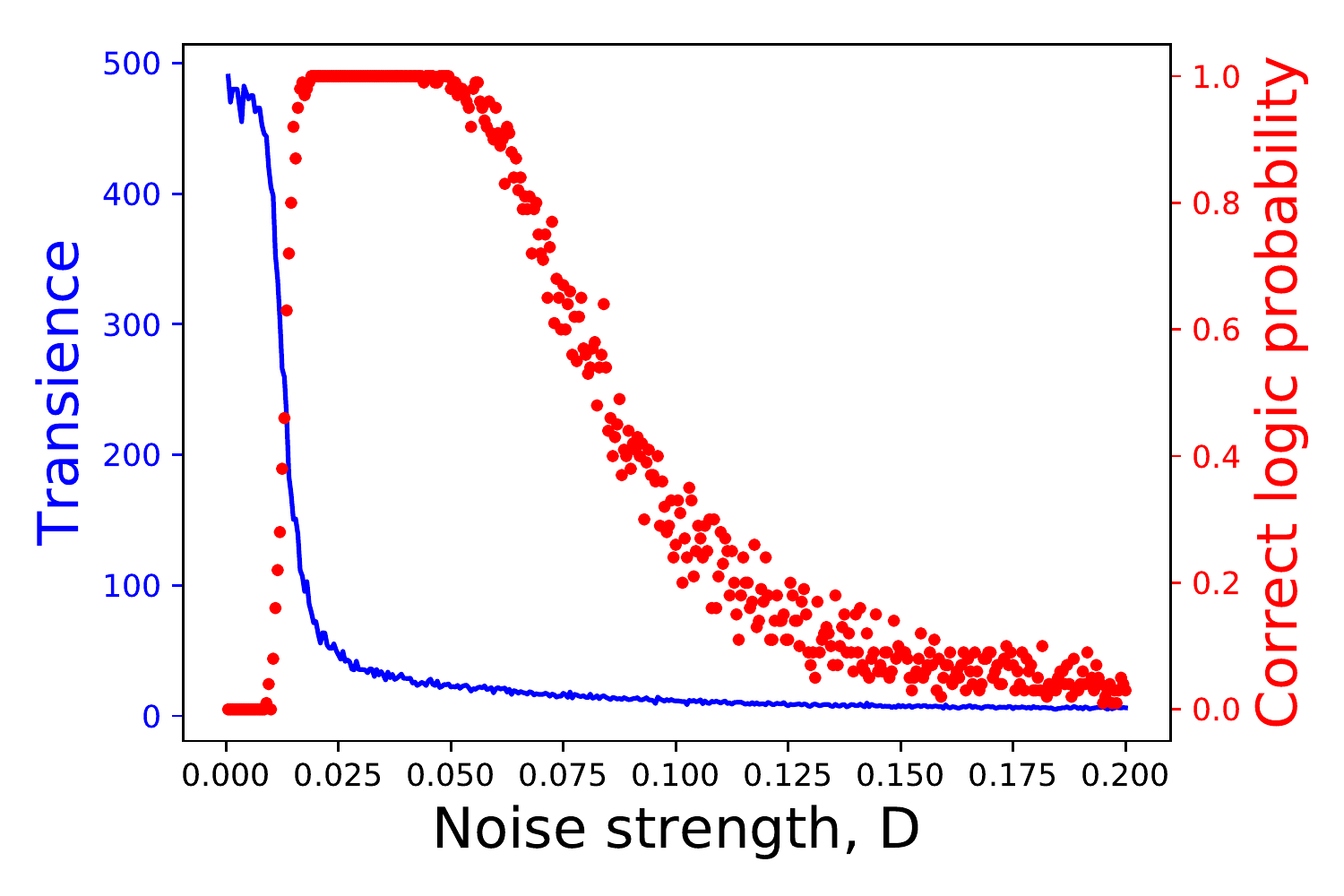}
\end{center}
\caption{Transience (averaged over a random stream of
    inputs) as a function of noise strength. Here transience is
    estimated from numerical simulations, and is the time taken to
    reach the barrier from a well, when the input switches necessitate
    a change in the output. It is shown in terms of the scaled time in
    Eqn.~3, where $1$ unit is $0.0000134$ $sec$. The system parameters
    are those given in Fig.~\ref{PNOR}.} 
  \label{transience}
\end{figure}

We also observed the reduction of latency with
  increasing noise.  This is evident in
  Fig.~\ref{transience}. Clearly, the system responds to inputs more
  rapidly when noise intensity is higher. So the desired hopping
  between wells happens faster under the influence of stronger
  noise. This is yet another feature where noise assists performance.\\

\bigskip

\noindent
{\bf Generalized Logical Stochastic Resonance with Input-Modulated Parameters}\\

\bigskip

\begin{figure}[ht]
\begin{center}
\includegraphics[width=1\linewidth,height=1\linewidth]{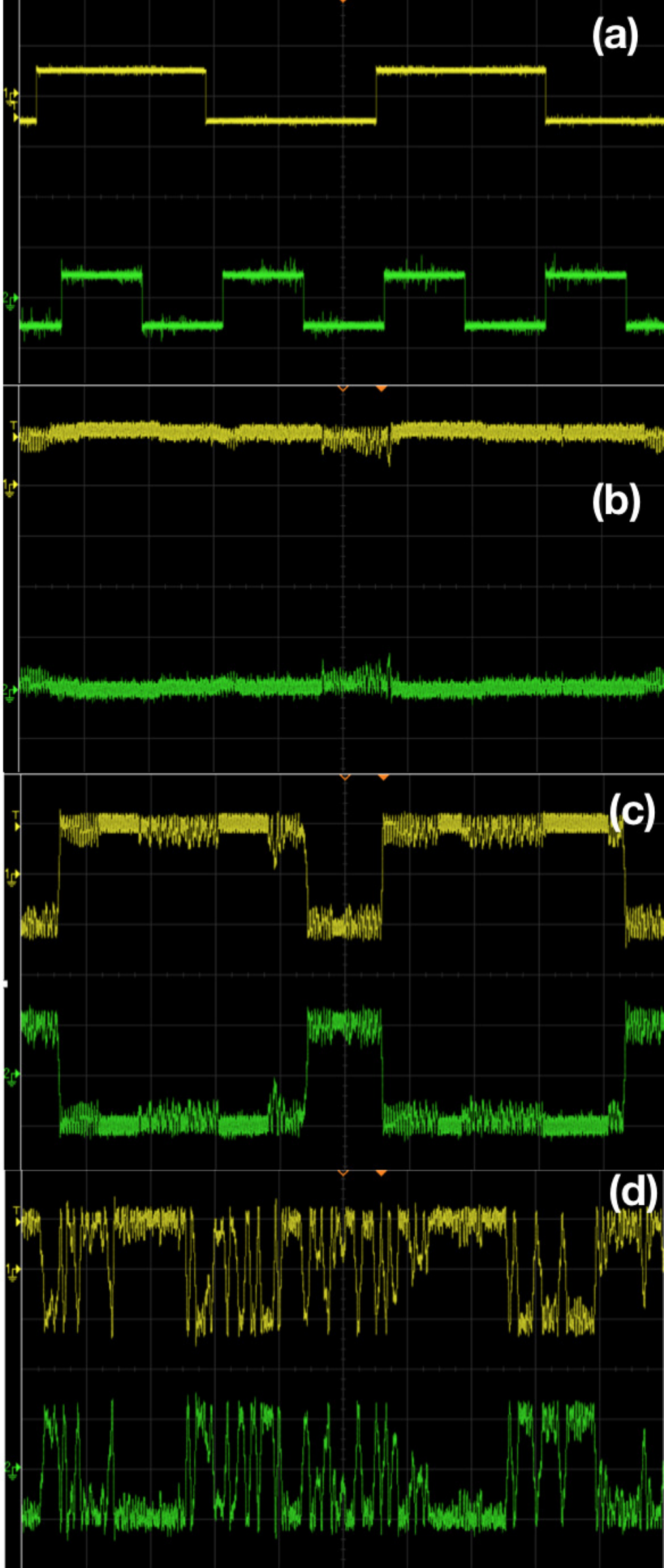}
\end{center}
\caption{Generalized Logical Stochastic Resonance with
    Input-Modulated Parameters: Results from electronic circuit
    experiments, with panel (a) showing stream of inputs $I_1$ and $I_2$
    (which take value $-0.5 \ V$ when logic input is $0$ and value
    $0.5 \ V$ when logic input is $1$). Panels (b) to (d) show the
    output voltage $V(t)$ (yellow) and $V_L (t)$ (green) for different
    noise strengths $D$: (i) $100 \ mV$, (ii) $1.0 \ V$ and (iii) $1.5
    \ V$. Clearly panel (c) depicts logical OR output (with $V(t) > 0$
    being logic output $1$, and $V(t) < 0$ being logic output
    $0$). The output voltage $V_L (t)$ (green) yields the
    complementary NOR logic gate response. The amplitude A of the
    sinusoidal signal is $100 \ mV$ and frequency is $8890 \ Hz$. The
    bias voltage $b$, is fixed as $20 \ mV$, and is modulated by the
    input signal streams as: $b(t) = b ( I_1 + I_2)$. A separate
    multiplier chip AD633 is used for modulation. }
\label{parametric}
\end{figure}

We demonstrate a further generalization of Logical Stochastic Resonance using input-modulated parameters, offering multiplicative perturbations to the system. Fig.~\ref{parametric} shows representative experimental results of this, for the system in Eqn.~\ref{mlc} with the input signal $I = I_1 + I_2$ modulating parameter $b$:
\begin{eqnarray}
\label{eqn_parametric}
\dot{x}&=&y - g(x) ,\\ \nonumber
\dot{y}&=& - a y - x + b (I_1 + I_2) + f(t) + D \eta (t) ,
\end{eqnarray}
where $\eta (t)$ is a zero-mean Gaussian noise with variance $1$, and parameter $D$ gives the noise strength.  In the circuit implementation (cf. Fig.~\ref{fig5}), $g(t)$ now corresponds to $b (I_1+I_2)+f(t) + D \eta(t)$. The important distinction with the system in Eqn.~\ref{mlc_noise} is that the {\em stream of inputs now modulate a parameter}. It is clearly evident that the system yields the appropriate logic output, as the input sets change, with the system switching as desired between different dynamical attractors bounded in distinct regions of phase space. Again, the logic output is obtained consistently in a window of moderate noise. This suggests that the scope of Logical Stochastic Resonance may be expanded to inputs-modulated parameters as well.

\bigskip 
\bigskip

\noindent
{\bf Conclusion}\\

\bigskip

The potential significance of the results obtained in this work are
two-fold.  The first is the proposal to implement fundamental logic
operations by exploiting the switching between chaotic attractors. The
underlying idea here is as follows: certain nonlinear systems can hop
between dynamical attractors occupying different regions of phase
space, under variation of parameters or initial states. We exploit
this feature to obtain reliable logic operations by explicitly
demonstrating the implementation of the fundamental NOR gate.  The
logic response can be morphed from NOR to NAND by a small change in
the bias parameter, and this flexibility lays the foundation for
general purpose reconfigurable circuitry \cite{fpga,ss2}. Further this
system offers the advantage that very low-amplitude inputs (of the
order of $10^{-3}$) yield highly amplified outputs (of the order of
$1$). The underlying reason for this is that small changes in the
system yield significantly different dynamical outcomes. Additionally,
different dynamical variables in the system yield complementary logic
operations in parallel.

The second signficant result here is a generalization
  of the concept of Logical Stochastic Resonance. We show how the idea
  of LSR, which has so far been realized using steady states, may be
  extended to more complex dynamical attractors. So the noise floor
  can aid the reliability of the logic operations even when the logic
  operation is based on switching between states more complex than
  fixed points, such as hopping between periodic cycles or chaotic
  attractors.  We also demonstrated that the generalized Logical
  Stochastic Resonance holds true for input-modulated parameters
  and multiplicative perturbations to the system. 

In summary, we have shown how hopping between
  dynamical attractors of different geometries can be exploited for
  the implementation of logic gates.  The ideas presented here,
  combining the research directions of Chaos Computing and Logical
  Stochastic Resonance, has potential to be realized in wide-ranging
  systems, and represents a new direction in exploiting chaotic systems
  to design computational devices.

\end{document}